\documentclass[preprint,pdf]{iucr}

\RequirePackage{graphicx}
\RequirePackage{amsmath}
\RequirePackage{amssymb}
\papertype{FA}
\paperlang{english}

\def\href#1{\relax}\let\foo\caption
\ifPDF
  \RequirePackage{hyperref}
  \PassOptionsToPackage{pdftex,bookmarksopen,bookmarksnumbered}{hyperref}
\fi
\let\caption\foo
\begin{document}         

\title{Using the singular value decomposition to extract 2D correlation functions from scattering patterns}
\shorttitle{2D correlation functions}

\cauthor[a]{Philipp}{Bender}{philipp.bender@uni.lu}{}
\author[b,c]{Dominika}{Z\'akutna}
\author[c]{Sabrina}{Disch}
\author[d,e]{Lourdes}{Marcano}
\author[f]{Diego}{Alba Venero}
\author[b]{Dirk}{Honecker}

\aff[a]{Physics and Materials Science Research Unit, University of Luxembourg, 162A~avenue de la Fa\"iencerie, L-1511 Luxembourg, \country{Grand Duchy of Luxembourg}}
\aff[b]{Large Scale Structures group, Institut Laue-Langevin, 71~avenue des Martyrs, F-38042 Grenoble, \country{France}}
\aff[c]{Department f\"ur Chemie, Universit\"at zu K\"oln, Luxemburger Strasse 116, D-50939 K\"oln, \country{Germany}}
\aff[d]{Helmholtz-Zentrum Berlin f\"ur Materialien und Energie,  Albert-Einstein-Strasse 15, D-12489 Berlin \country{Germany}}
\aff[e]{Departamento Electricidad y Electrónica, Universidad del País Vasco – UPV/EHU, E-48940 Leioa, \country{Spain}}
\aff[f]{ISIS Neutron and Muon Facility, Rutherford Appleton Laboratory,
Chilton, OX11 0QX, \country{United Kingdom}}

\keyword{small-angle scattering}
\keyword{correlation function}
\keyword{two-dimensional Fourier transform}
\keyword{anisotropic structures}
\keyword{nanoparticles}
\keyword{singular value decomposition}
\keyword{noise filtering}

\maketitle                       

\begin{synopsis}
We apply the truncated singular value decomposition (TSVD) to extract the underlying 2D correlation functions from small-angle scattering patterns.
\end{synopsis}

\begin{abstract}
We apply the truncated singular value decomposition (TSVD) to extract the underlying 2D correlation functions from small-angle scattering patterns.
We test the approach by transforming the simulated data of ellipsoidal particles and show that also in case of anisotropic patterns (i.e. aligned ellipsoids) the derived correlation functions correspond to the theoretically predicted profiles.
Furthermore, we use the TSVD to analyze the small-angle x-ray scattering patterns of colloidal dispersions of hematite spindles and magnetotactic bacteria in presence of magnetic fields, to verify that this approach can be applied to extract model-free the scattering profiles of anisotropic scatterers from noisy data.
\end{abstract}

\section{Introduction}

Small-angle angle scattering of x-rays (SAXS) or neutrons (SANS) is routinely used to investigate the structural properties of various materials, such as nanostructered bulk samples \cite{fritz2013two,fritz2015interpretation,michels2014magnetic} or nanoparticle ensembles \cite{reufer2010morphology,markert2011small,hoffelner2015directing,nack2018hindered}.
In many cases, the detected 2D scattering patterns $I(\textbf{q})$ (here $I$ is the scattering intensity and $\textbf{q}$ the scattering vector on the 2D detector plane), are isotropic and data analysis is performed on the 1D azimuthal average $I(q)$, by either fitting it in reciprocal space with model functions or by transforming $I(q)$ to real-space correlation functions $C(r)$ \textit{via} Fourier transforms (FT) \cite{pedersen1997analysis}.
For the latter, a direct FT can be used \cite{michels2003range} or, alternatively, an indirect FT \cite{glatter1977new,hansen2000bayesian,bender2017structural}.
Often the pair distance distribution function $P(r)=C(r)r^2$ is determined, which gives direct information about the average shape, size and structure of the scattering units \cite{glatter1979interpretation,svergun2003small}.
In case of anisotropic nanostructures, the pattern $I(\textbf{q})$ can be anisotropic, provided the nanostructures are preferentially aligned along a certain director or texture axis \cite{van2004orientation,reufer2010morphology,markert2011small,fritz2013two,hoffelner2015directing,nack2018hindered}.
Also in this case data analysis can be either performed by adjusting the whole scattering pattern in reciprocal space \cite{alves2017calculation} or by transforming the pattern \textit{via} 2D FTs to the real-space 2D correlation functions \cite{fritz2013two,fritz2015interpretation,mettus2015small}.
For the indirect 2D FT proposed in \citeasnoun{fritz2013two}, regularization functionals are appended to the system matrix, which is necessary to obtain smooth profiles, but which can increase the computation times significantly. 
Here, we propose a similar but faster method, namely the truncated singular value decomposition (TSVD), to extract the underlying 2D correlation functions.
We propose that this method could be used for live, model-free data analysis, e.g. in capillary flow devices \cite{nielsen2012high}, time-resolved nanoparticle-oscillation \cite{bender2015excitation} or rheo-SAXS/SANS experiments \cite{panine2003combined,calabrese2016optimized}, and microfluidics \cite{stehle2013small,poulos2016microfluidic,adamo2018droplet}, where the noise reduction of the SVD helps to enhance the signal.

\section{Methods}
\subsection{Singular value decomposition SVD}

In polar coordinates the 2D scattering intensity $I(q_y,q_z)=I(q,\Theta)$ (the beam is along the $x$-direction, $\mathbf{k}||\mathbf{e_x}$) is a function of the magnitude of the scattering vector $|\mathbf{q}|$ and the angle $\Theta$. 
With $i=N$ unique data points (i.e. pixels of the detector for a monochromatic source) it can be written
\begin{equation}\label{Eq3}
I(q_i,\Theta_i)=\sum_{k=1}^K A_{ik}P(r_k,\varphi_k),
\end{equation}
where the angle $\varphi$ specifies the orientation of $\mathbf{r}$ in real space in the $yz$ plane.
Here, the extracted 2D distribution function is $P(r,\varphi)=C(r,\varphi)r$ \cite{fritz2013two}, and the matrix $\mathbf{A}$ has the elements \cite{mettus2015small} 
\begin{equation}\label{Eq4}
A_{ik}=\mathrm{cos}\left(q_ir_k\mathrm{cos}\left(\Theta_i-\varphi_k\right)\right)\Delta r_k\Delta\varphi_k.
\end{equation}
Here, we usually assume a linear spacing for the pre-determined $r$- and $\varphi$-vectors (i.e. $\Delta r$ and $\Delta\varphi=\mathrm{const}$).
By minimizing the deviation (i.e. performing a least-square fit)
\begin{equation}\label{Eq5}
\chi^2=\sum_{i=1}^N\frac{\left(I(q_i,\Theta_i)-\sum_{k=1}^K A_{ik}P(r_k,\varphi_k)\right)^2}{\sigma_i^2}
\end{equation}
the 2D correlation function $P(r,\varphi)=C(r,\varphi)r$ can be determined. 
In case of significant measurement uncertainties $\sigma_i$, however, this distribution will exhibit strong, unphysical oscillations.
While performing the indirect Fourier transform, regularization matrices are appended (e.g. a Tikhonov regularization can be applied) to force smooth distributions \cite{fritz2013two}, and various approaches exist to estimate the optimal value for the regularization parameter \cite{glatter1977new,hansen2000bayesian}.

With the SVD the decomposition of the system matrix $\mathbf{A}=\mathbf{U}\mathbf{S}\mathbf{V}^T$ is performed, where $\mathbf{U}$ and $\mathbf{V}$ are orthogonal $N\times N$ and $K\times K$ matrices, respectively, and $\mathbf{S}$ is a $N\times K$ matrix whose main diagonal elements $s_i\equiv s_{ii}$ are the singular values (all other values are zeros).
The solution $P(r,\varphi)$ of the functional $\mathbf{A}P(r,\varphi)=I(q,\Theta)$ is
\begin{equation}\label{Eq6}
P(r,\varphi)=\mathbf{V}\mathbf{S}^{+}\mathbf{U}^TI(q,\Theta),
\end{equation}
where $\mathbf{S}^{+}$ has the reciprocal values $1/s_i$ in its diagonal, and otherwise zeros.
Using the full-rank matrices for the reconstruction of $P(r,\varphi)$ according to Eq.\,\ref{Eq6} should result in the same distribution as minimizing Eq.\,\ref{Eq5}.
In particular, very small $s_i$ values associated to noise result in strong oscillations of the reconstructed distribution.
Thus, a smoothing of $P(r,\varphi)$ can be achieved by reducing the number of singular values $s_i$ that are considered for $\mathbf{S}^{+}$ \cite{0022-3727-33-4-303} (i.e. singular values which are smaller than a certain threshold are set to zero).
Alternatively, a TSVD can be performed \cite{hansen1987truncatedsvd,viereck2019multi}, which further reduces the computation time compared to a SVD of the full-ranked matrices.

We will use the TSVD to analyze first the simulated scattering patterns of an ensemble of prolate ellipsoids and later the measured SAXS patterns of a colloidal dispersion of hematite spindles and of magnetotactic bacteria.

\subsection{Samples}
Details regarding the synthesis and characterization of the hematite spindles can be found in \cite{C8NR09583C}. 
Hematite nanospindles are weakly ferromagnetic and known to orient with their main axis perpendicular to an applied magnetic field \cite{reufer2010morphology,markert2011small,roeder2015magnetic,hoffelner2015directing,nack2018hindered}. 
According to electron microscopy, the sample studied here has a length of the short spindle axes of around 56\,nm and of the long axis of around 375\,nm.
The SAXS measurements from \cite{C8NR09583C}, which we also analyze here, were performed at the ID02 instrument at the ESRF \cite{narayanan2018multipurpose} on a dilute, i.e. non-interacting, dispersion of the nanospindles. A static homogeneous magnetic field up to 520mT was applied with an electromagnet in horizontal direction perpendicular to the incoming beam.

A detailed study of the magnetotactic bacteria is presented in \cite{orue2018configuration}.
These bacteria contain on average about 15-20 iron oxide nanoparticles with diameters of around 40-50\,nm, which are typically arranged in nearly linear chains.
Here we analyze a SAXS measurement of the same colloidal dipsersion of bacteria as presented in \citeasnoun{orue2018configuration}. 
This measurement was performed with a Xenocs Nano-InXider.
The colloidal dispersion was filled into a quartz glass capillary and the bacteria aligned perpendicular to the incoming beam by positioning a small permanent magnet next to the capillary.
With this setup only one full quadrant of the pattern is detected but assuming symmetry the total $2\pi$ pattern can be easily reconstructed.

\section{Results}
\subsection{Simulated data}

\begin{figure}\label{Fig1}
\centering
\includegraphics[width=1\columnwidth]{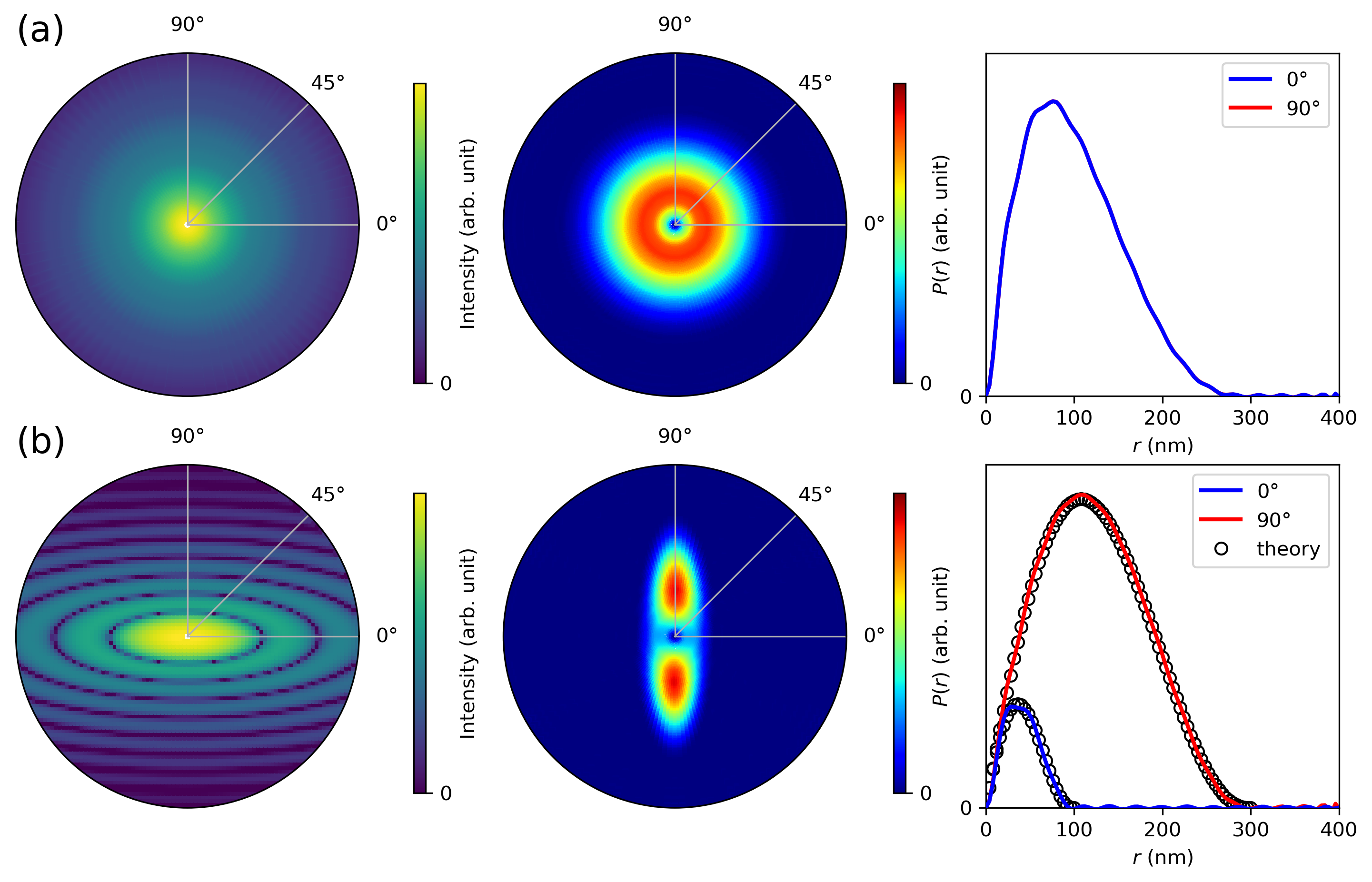}
\caption{Determination of the 2D correlation functions by a SVD for a monodisperse ensemble of prolate rotational ellipsoids with an equatorial radius of 50\,nm and a polar radius of 150\,nm. (a) Ellipsoids are randomly oriented (isotropic ensemble). (b) Ellipsoids are aligned with their polar axis along $\Theta=90^\circ$ (anisotropic ensemble). \textit{Left column:} Polar plot of the scattering pattern in logarithmic scale ($q_\mathrm{max}=0.2\,\mathrm{nm^{-1}}$). \textit{Middle column:} Extracted 2D correlation function $P(r,\varphi)$ ($r_\mathrm{max}=400\,\mathrm{nm}$). \textit{Right column:} Correlation function along $\varphi=0^\circ$ and $90^\circ$, and in (b) additionally the calculated cross sections for spheres with diameters of 100\,nm and 300\,nm, respectively.}
\end{figure}

\begin{figure}\label{Fig2}
\centering
\includegraphics[width=0.8\columnwidth]{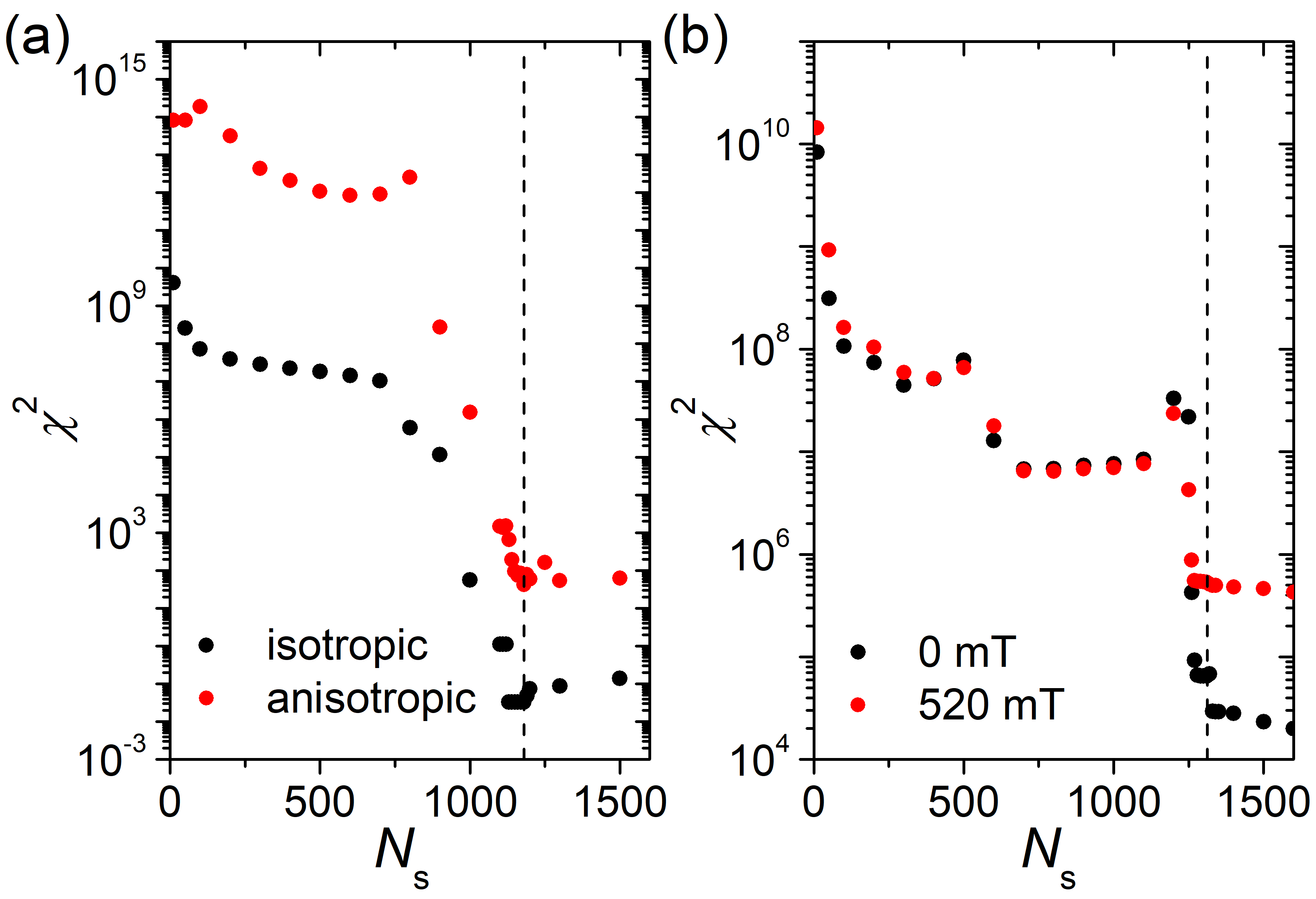}
\caption{Error $\chi^2$ (Eq.\,\ref{Eq5}) for the fit of (a) the simulated data (Fig.\,\ref{Fig1}), and (b) the experimental data of the spindles (Fig.\,\ref{Fig3}) at two different fields, with
$N_\mathrm{s}$ being the number of singular values used for the reconstruction.}
\end{figure}

To test the TSVD, we first analyze the calculated scattering patterns of rotational ellipsoids with an equatorial radius of 50\,nm and a polar radius of 150\,nm (\textit{Left column} in Figs.\,\ref{Fig1}(a) and (b)).
In Fig.\,\ref{Fig1}(a) (upper row) we analyze the isotropic ensemble and in Fig.\,\ref{Fig1}(b) (lower row) the anisotropic ensemble.
The simulated scattering patterns had $N=30480$ data points, and we calculated the system matrix $\mathbf{A}$ (30480$\times$10224 matrix) for 101 $r$-values ($0 - 400\,\mathrm{nm}$, 4\,nm-steps) and 144 $\varphi$-values ($2.5 - 360^{\circ}$, $2.5^{\circ}$-steps).
We then varied the number of singular values $N_\mathrm{s}$ considered for the reconstruction.
Usually, $N_\mathrm{s}$ is much smaller than the dimensions of system matrix, and thus the TSVD significantly reduces the computational complexity compared for example to the classical IFT approaches.
As shown in Fig.\,\ref{Fig2}(a), the total error $\chi^2$ is significantly reduced by increasing $N_s$ and exhibits a minimum at around $N_\mathrm{s}=1180$, which is marked by the dashed line (here we used $\sigma_i^2=I(q_i,\Theta_i)$ for the calculation of $\chi^2$, Eq.\,\ref{Eq5}).
Further increasing $N_\mathrm{s}$ does not minimize the deviations more, but will generate strong oscillations due to the restricted $q$-range \cite{qiu2004pdfgetx2}.
In Fig.\,\ref{Fig1} we plot the reconstructed distributions for $N_\mathrm{s}=1180$, and in case of the anisotropic ensemble compare them to the theoretically expected profiles $C(r)r$ \cite{fritz2015interpretation}, which we calculated for spheres with diameters of 100\,nm and 300\,nm, respectively. 
The good agreement shows that the TSVD can be readily used for a fast determination of the underlying 2D correlation functions from 2D scattering patterns, at least in case of smooth data.

\subsection{Experimental data}
\subsubsection{Hematite spindles}

\begin{figure}\label{Fig3}
\centering
\includegraphics[width=1\columnwidth]{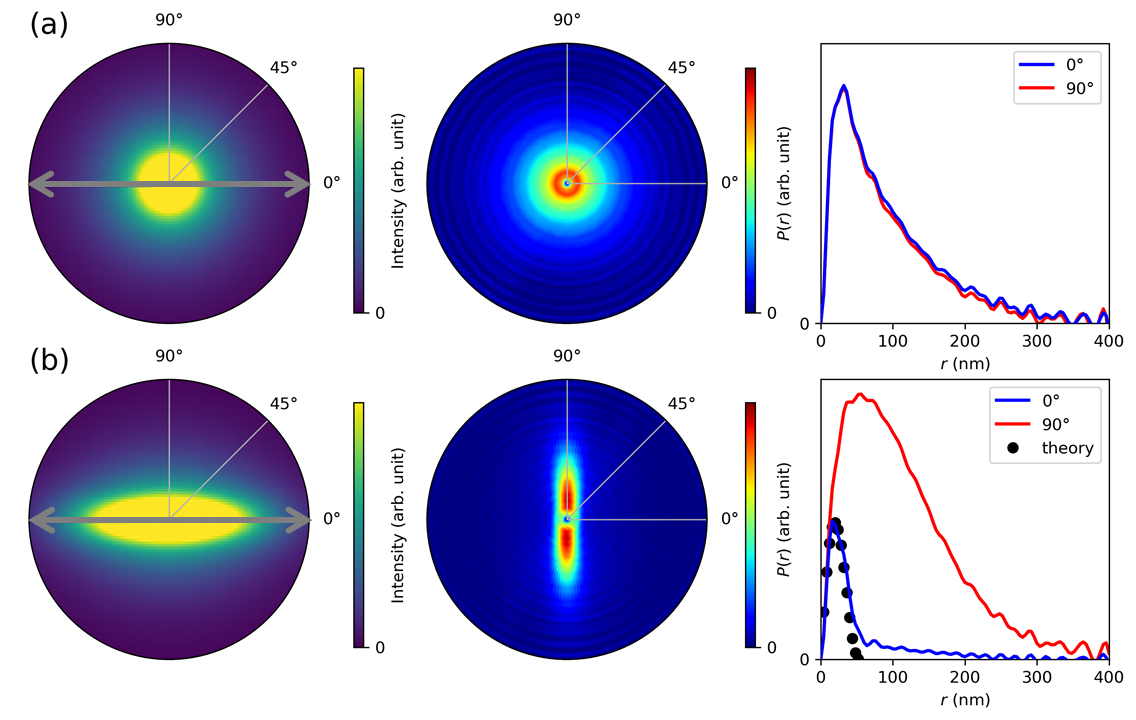}
\caption{Analysis of the 2D SAXS patterns of the colloidal dispersion of hematite spindles; the magnetic field was applied in horizontal direction ($\Theta=0^\circ$) with a field strength of (a) 0\,mT and (b) 520\,mT.
\textit{Left column:} Polar plot of the scattering pattern in logarithmic scale ($q_\mathrm{max}=0.1\,\mathrm{nm^{-1}}$). \textit{Middle column:} Extracted 2D correlation function $P(r,\varphi)$ ($r_\mathrm{max}=400\,\mathrm{nm}$). \textit{Right column:} Correlation function along $\varphi=0^\circ$ and $90^\circ$, and the theoretically expected cross sections for a sphere with a diameter of 56\,nm (black dots).}
\end{figure}

In Fig.\,\ref{Fig3} we show the analysis results of the field-dependent scattering patterns of the colloidal dispersion of hematite spindles.
In the upper row (Fig.\,\ref{Fig3}(a)) we display the results for 0\,mT and in the lower row (Fig.\,\ref{Fig3}(b)) the results for $\mu_0H=520\,\mathrm{mT}$.
For both, the isotropic and the anisotropic pattern, the SVD was performed for 101 $r$-values ($0 - 400\,\mathrm{nm}$) and 144 $\varphi$-values ($2.5 - 360^{\circ}$, $2.5^{\circ}$-steps).
The correlation functions plotted in Fig.\,\ref{Fig3} were determined for $N_s=1270$, for which the total error was already close to the minimum (Fig.\,\ref{Fig2}(b)).
For both cases the extracted correlation functions exhibit some oscillations, but which increases significantly with increasing $N_s$ (i.e. smaller singular values are associated to signal noise).

At 0\,mT, the anisotropic particles are randomly oriented within the viscous matrix (water), and thus the scattering pattern and the derived correlation function are both isotropic.
The extracted profile has a pronounced peak at around 50\,nm and a long tail, and hence corresponds well to the expected profile of randomly-oriented, shape-anisotropic nanoparticles \cite{svergun2003small}.

At 520\,mT on the other hand, the particle moments are mostly oriented in field direction.
The dominance of the magnetocrystalline anisotropy leads to a net magnetic moment in the basal plane of the hematite spindles, i.e. perpendicular to the long particle axis \cite{reufer2010morphology,hoffelner2015directing,nack2018hindered}.
Therefore, the long spindle axes are oriented within the plane perpendicular to the field direction, which explains the anisotropy of both the scattering pattern and the extracted correlation function.
Along $\varphi=90^\circ$ (\textit{right column} of Fig.\,\ref{Fig3}(b)) the profile indicates a maximum length of the spindles of around 400\,nm, whereas the profile along $\varphi=0^\circ$ agrees well with the expected correlation function for spindles with an equatorial radius of 56\,nm.
Thus, the extracted 2D correlation function is in good agreement with our previous results  \cite{C8NR09583C}.

\subsubsection{Magnetotactic bacteria}
\begin{figure}\label{Fig4}
\centering
\includegraphics[width=1\columnwidth]{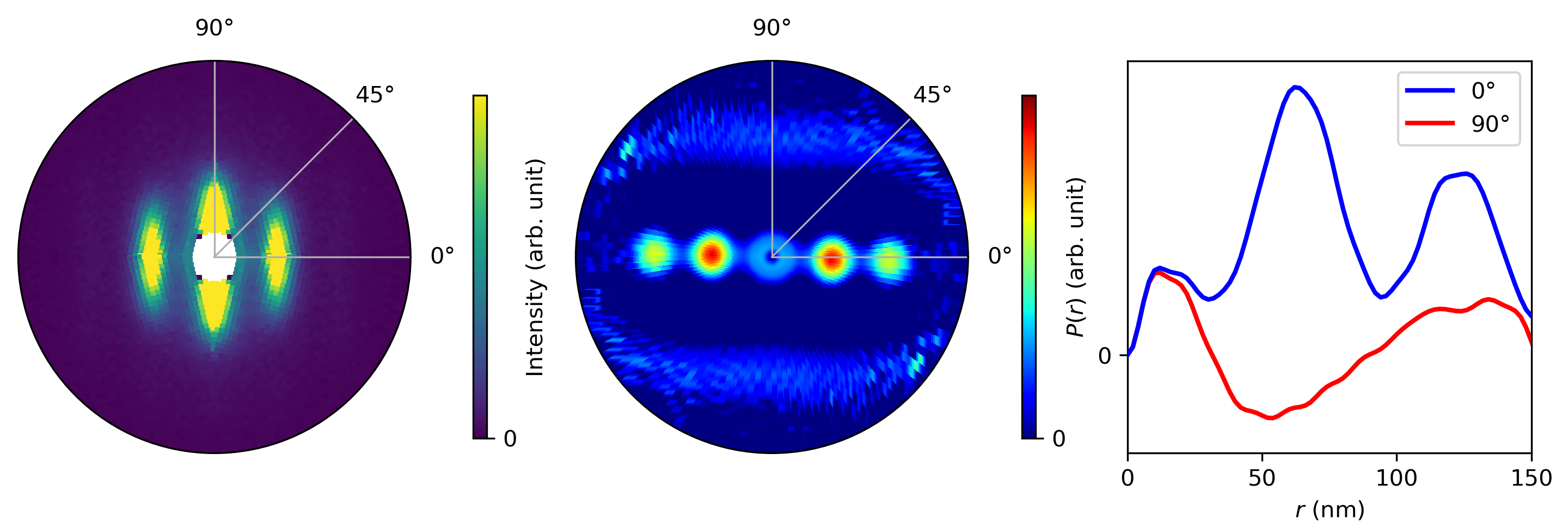}
\caption{Analysis of the 2D SAXS pattern of the colloidal dispersion of magnetotactic bacteria; the magnetic field was applied in horizontal direction ($\Theta=0^\circ$).
\textit{Left column:} Polar plot of the scattering pattern in logarithmic scale ($q_\mathrm{max}=0.5\,\mathrm{nm^{-1}}$). 
\textit{Middle column:} Extracted 2D correlation function $P(r,\varphi)$ ($r_\mathrm{max}=150\,\mathrm{nm}$).
\textit{Right column:} Correlation function along $\varphi=0^\circ$ and $90^\circ$.}
\end{figure}

In Fig.\,\ref{Fig4} we show the highly anisotropic SAXS pattern of the aligned magnetotactic bacteria along the horizontal field direction.
The extracted correlation function was determined, as before, for 101 $r$-values ($0 - 200\,\mathrm{nm}$) and 144 $\varphi$-values ($2.5 - 360^{\circ}$, $2.5^{\circ}$-steps).
The function plotted in Fig.\,\ref{Fig4} was determined for $N_\mathrm{s}=1000$, which shows that the particles are preferentially spheres (radially symmetric spots), and that the bacteria and thus the magnetosome chains are well aligned along the applied field showing no spreading of the spots in the azimuthal direction.
From the profile along $\varphi=0^\circ$ (\textit{right column} of Fig.\,\ref{Fig4}) we can estimate the average center-to-center distance between the particles in the chain from the peak-to-peak distance to be around 50\,nm and the particle size from the peak width to be around 40\,nm, which agrees well with previous results \cite{orue2018configuration}.
Along $\varphi=90^\circ$ we only see the cross section of a single particle. 
However, for $r>40$\,nm the profile oscillates around zero.
This indicates a variation of the scattering length density profile, and which we attribute to a lower scattering length density in the vicinity of the particle (i.e. the bacteria) compared to the surrounding matrix (i.e. the water).

\section{Conclusions}

Here, we introduce the truncated singular value decomposition (TSVD) to extract the denoised two-dimensional correlation functions from small-angle scattering patterns.
First, we analyze simulated data of ellipsoidal particles and show that the derived correlation functions correspond well to the theoretically predicted profiles.
Furthermore, we use this method to succesfully extract the underlying 2D correlation functions from experimental SAXS patterns of colloidal dispersions of hematite spindles and magnetotactic bacteria.
We emphasize that this method is a fast and easy way to obtain model-free information about the structural properties of anisotropic scatterers, which we propose for example for the analysis of live data.

\ack{P. Bender acknowledges financial support from the National Research Fund of Luxembourg (CORE SANS4NCC grants), L. Marcano acknowledges the Basque Government for her fellowship (POS{\_}2018{\_}1{\_}0070) and S. Disch acknowledges financial support from the German Research Foundation (DFG: DI 1788/2-1).
The SAXS experiments on the hematite spindles were performed on beamline ID02 at the European Synchrotron Radiation Facility (ESRF), Grenoble, France, and we are grateful to  Sylvain Prévost for providing assistance.
We thank the group of Magnetism and Magnetic Materials at the Universidad del País Vasco for preparing the magnetotcatic bacteria and the Science and Technology Facilities Council (STFC) for access to the SAXS kit at the Materials Characterisation Laboratory to perform the SAXS measurement.}

\newpage

\referencelist[PBenderBib]

\end{document}